\pdfoutput=1 
\documentclass[sigconf, nonacm]{acmart}
\acmConference[??? 2023]{37th IEEE/ACM ???}{October 10--14, 2022???}{Michigan, United States???}
\setcopyright{acmcopyright}
\copyrightyear{2022}
\acmYear{2022}
\acmDOI{TBA}

\AtBeginDocument{%
  \providecommand\BibTeX{{%
    \normalfont B\kern-0.5em{\scshape i\kern-0.25em b}\kern-0.8em\TeX}}}

\setcopyright{acmcopyright}
\copyrightyear{2018}
\acmYear{2018}
\acmDOI{10.1145/1122445.1122456}

\acmConference[Woodstock '18]{Woodstock '18: ACM Symposium on Neural
  Gaze Detection}{June 03--05, 2018}{Woodstock, NY}
\acmBooktitle{Woodstock '18: ACM Symposium on Neural Gaze Detection,
  June 03--05, 2018, Woodstock, NY}
\acmPrice{15.00}
\acmISBN{978-1-4503-XXXX-X/18/06}

\usepackage[pagewise]{lineno}
 \usepackage{amsmath}

\usepackage{algorithm}
\usepackage{algorithmic}
\usepackage{amsmath}
\usepackage{CJKutf8}
\usepackage{graphicx}
\usepackage{float}
\usepackage{multirow}
\usepackage{amsfonts}
\usepackage{galois}
\usepackage{amssymb}
\usepackage{enumerate}
\usepackage{subfigure}
\usepackage{threeparttable}

\usepackage{colortbl}
\usepackage{color, xcolor}
\usepackage{diagbox}
\usepackage{tcolorbox}
\usepackage{threeparttable}
\usepackage{bm}

\usepackage{mathrsfs}
\usepackage{makecell}
\usepackage{soul}
\usepackage{listings}

\definecolor{dkgreen}{rgb}{0,0.6,0}
\definecolor{gray}{rgb}{0.5,0.5,0.5}
\definecolor{mauve}{rgb}{0.58,0,0.82}
\newcommand{\syrevised}[1]{\textcolor{black}{#1}}
\newcommand{\tbd}[1]{\textcolor{black}{#1}}

\newcommand{\vtwo}[1]{\textcolor{black}{#1}}
\newcommand{\tbdtwo}[1]{\textcolor{black}{#1}}

\newcommand{\grz}[1]{\textcolor{black}{#1}}
\newcommand{\sy}[1]{\textcolor{black}{#1}}

\begin{document}

\title{A novel failure indexing approach with run-time values of program variables}


\author{Yi Song}
\affiliation{%
	\institution{School of Computer Science,\quad Wuhan University}
	\city{Wuhan}
	\country{China}}
\email{yisong@whu.edu.cn}

\author{Xihao Zhang}
\authornote{Co-first author.}
\affiliation{%
	\institution{School of Computer Science,\quad Wuhan University}
	\city{Wuhan}
	\country{China}}
\email{zhangxihao@whu.edu.cn}

\author{Xiaoyuan Xie}
\authornote{Corresponding author.}
\affiliation{%
	\institution{School of Computer Science,\quad Wuhan University}
	\city{Wuhan}
	\country{China}}
\email{xxie@whu.edu.cn}

\author{Quanming Liu}
\affiliation{%
	\institution{School of Computer Science,\quad Wuhan University}
	\city{Wuhan}
	\country{China}}
\email{liuquanming@whu.edu.cn}

\author{Ruizhi Gao}
\affiliation{%
	\institution{Sonos Inc.}
	\city{Santa Barbara}
	\country{USA}}
\email{youtianzui.nju@gmail.com}

\author{Chenliang Xing}
\affiliation{%
	\institution{School of Computer Science,\quad Wuhan University}
	\city{Wuhan}
	\country{China}}
\email{xingchenliang@whu.edu.cn}

\renewcommand{\shortauthors}{Yi Song, Xihao Zhang, Xiaoyuan Xie, Quanming Liu, Ruizhi Gao and Chenliang Xing}

\begin{abstract}
	Failures with different root causes can disturb multi-fault localization significantly, therefore, dividing failures into distinct groups according to the responsible faults is \grz{highly important}. In such a \textbf{failure indexing} task, the crux lies in \vtwo{the failure proximity, which involves} two points, i.e., how to effectively represent failures (e.g., extract the signature of failures) and how to properly measure the distance between the proxies for those failures. Existing studies have proposed a variety of \vtwo{failure proximities}. \vtwo{The prevalent} of them extract signatures of failures \vtwo{from execution coverage or suspiciousness ranking lists, and accordingly employ the Euclid or the Kendall tau distances. However, such strategies may not properly reflect the essential characteristics of failures, thus  resulting in unsatisfactory effectiveness.} In this paper, we propose a new failure proximity, \vtwo{namely, program variable-based failure proximity, and based on which present a novel failure indexing approach.} Specifically, the proposed approach utilizes the run-time values of program variables to represent failures, and designs a set of rules to measure the similarity between them. Experimental results demonstrate the competitiveness of the proposed approach: it can achieve \vtwo{44.12\%} and \vtwo{27.59\%} improvements in faults number estimation, as well as \vtwo{47.30\%} and \vtwo{26.93\%} improvements in clustering effectiveness, compared with the state-of-the-art technique in this field, in simulated and real-world environments, respectively.
	
\end{abstract}



\settopmatter{printfolios=true}

\maketitle

\section{Introduction}
\label{sect:intro}

Nowadays, the mainstream research direction in fault localization focuses on single-fault scenarios. However, with the increasing scale of software systems in modern development, such an assumption is being unrealistic in practice~\cite{digiuseppe2015fault[zxh21],gao2018research[zxh22],jones2002visualization[zxh23],wang2008software[zxh24]}. When programs contain more than one fault, many challenges could occur. For example, decrease of fault localization effectiveness incurred by the \vtwo{interference} between/among multiple faults~\cite{feng2018empirical[zxh44]}. Considering such \grz{an issue}~\cite{digiuseppe2011influence[6],digiuseppe2011fault[zxh38]}, developers prefer to employ \textbf{\emph{parallel debugging}}, where all \emph{failures}\footnote{Also known as \emph{failed test case} in the context of dynamic testing. We use these two terms interchangeably hereafter.} are divided into several disjoint groups according to their root causes~\cite{pei2021dynamic[zxh25],zakari2019community[zxh26],digiuseppe2012concept[zxh27],steimann2012improving[zxh28],golagha2019failure[10]}. This division\footnote{In the current field of parallel debugging, \emph{clustering} is typically utilized for such \emph{division}. Thus we use these two terms interchangeably hereafter.} process aims at two points, namely, 1) having the number of generated groups equal to the number of faults (i.e., \emph{correct faults number estimation}), and 2) failed test cases in the same group (referred to as fault-focused group) are triggered by the same fault\vtwo{, and vice versa} (i.e., \emph{high clustering effectiveness}). As such, each developer can be allocated to a fault-focused group and thus localize the corresponding fault independently and simultaneously.

The effectiveness of clustering determines the labor and time costs, as well as the performance, of parallel debugging. Specifically, if the number of faults is correctly predicted, and the clustering output is enough to distinguish different faults, a hunk of multi-fault localization task can be properly decomposed to several sub-single-fault localization tasks, resulting in faster and more effective delivering of failure-free software. Otherwise, in case of over-division (i.e., the predicted number of faults exceeds the truth), redundant developers will be expropriated for the debugging, resulting in labor waste. And in case of under-division (i.e., the predicted number of faults is less than the truth), more than one iteration of debugging is needed. \vtwo{Moreover, prior studies have proven that the higher the accuracy of clustering, the better the performance of parallel debugging, and vice versa~\cite{wu2020fatoc[zxh45], li2019empirical[41]}.} Thus, properly dividing failures (a.k.a \textbf{\emph{failure indexing}}) is the core of parallel debugging.

It is well-recognized that there are three essential factors in failure indexing~\cite{liu2008systematic[1]}: the fingerprinting function, the distance metric, and the clustering algorithm. The fingerprinting function is responsible for failure representation (e.g., by extracting signatures of failures), which products the proxy for failed test cases. The distance metric measures the similarity between proxies for failures, which relies on the form of the signature extracted by the fingerprinting function. The clustering algorithm builds linkages between failures and underlying faults incorporating the distance information. Previous research has concluded that there is no clustering technique that is universally applicable in uncovering the variety of structures present in multidimensional data sets~\cite{jain1999data[2]}. Therefore, the core of failure indexing lies in defining a proper fingerprinting function and designing a tailored distance metric (these two components are called \textbf{\emph{failure proximity}}).

A large number of researchers have dedicated their effort to exploring this topic~\cite{liu2008systematic[1],tu2016code[zxh1],liu2007indexing[zxh2],gao2019mseer[8],liu2006failure[zxh3]}, yielding a variety of failure proximities. Among them, the code coverage (CC)-based and the statistical debugging (SD)-based failure proximities \vtwo{are most widespread and commonly-used, previous studies have demonstrated their sophistication or advancement}~\cite{cao2017multiple[zxh29],wang2014weighted[zxh30],yu2015does[zxh31]}. \vtwo{As for the CC-based strategy, it employs binary coverage or execution frequency as its fingerprinting function, and typically uses the Euclid distance as its distance metric. And as for the SD-based strategy, it produces a suspiciousness ranking list of program entities as the proxy for a failure by incorporating fault localization techniques, which are generally SBFL (Spectrum-based Fault Localization) ones at the statement granularity~\cite{xie2013theoretical[31], jones2005empirical[4], xie2010isolating[35]}, and typically employs the Kendall tau distance, the Euclid distance, the Jaccard distance, etc. to measure the similarity between the proxies}~\cite{jaccard1912distribution[zxh6],kendall1948rank[zxh7]}. \vtwo{However, both of the two strategies have drawbacks. Specifically, if a fault triggers failures in different ways, \grz{i.e., failed test cases with the same root cause have different coverage,} effectiveness of the CC-based failure proximity could be threatened. Despite the integration of SBFL techniques~\cite{yoo2017human[32]}, the SD-based tactic also relies only on program coverage, resulting in unsatisfactory \vtwo{failure indexing} effectiveness when the coverage of the failures having distinct root causes is identical. \grz{One of the biggest drawbacks of using coverage to represent failures can be partly found in the PIE (Propagation, Infection, and Execution) model~\cite{voas1992pie[44]}. The PIE model thinks that a failure can be detected only if the fault infects the program's internal state, while coverage is hard to explore the internal state in depth during the program execution, thus cannot extract the signature of failures in deep insight.} Therefore, neither of CC and SD strategies is sufficient for serving as an effective failure representer (we use a motivating example to illustrate this point in Section~\ref{sect:motivation}).}

\vtwo{In our opinion, program internal dataflows could play a role of failure distinguisher when using coverage is in unsatisfactory effectiveness. Thus, in this paper, we propose the program variable-based failure proximity, which represents failures by run-time program variable information (i.e., the run-time values of program variables) and measures the distance based on the characteristic of such variable information. According to the intuition of the program variable-based failure proximity, we present a novel failure indexing approach to representing and clustering failed test cases.} For the fingerprinting function, the proposed approach first uses an SBFL technique to determine several riskiest program statements as breakpoints, and then collects run-time program variable information at the pre-set breakpoints during executing a failed test case. For the distance metric, the proposed approach designs a two-level framework to measure the similarity between a pair of program variable information that serves as proxies for failures. 

For the evaluation, we download four projects from SIR~\cite{do2005supporting[12]}, $flex$, $grep$, $gzip$, and $sed$, and inject mutated faults into clean programs, to generate 600 simulated faulty versions that contain one, two, three, four, or five faults (referred to as 1-bug, 2-bug, 3-bug, 4-bug, and 5-bug faulty version, respectively). We also gather \vtwo{100} real-world faulty versions containing 1\textasciitilde5 faults, from five projects in Defects4J~\cite{just2014defects4j[13]}, $Chart$, $Closure$, $Lang$, $Math$, and $Time$. Experimental results indicate that the proposed approach exceeds the state-of-the-art failure indexing technique significantly, with increases of \vtwo{44.12\%} and \vtwo{27.59\%} regarding faults number estimation, as well as increases of \vtwo{47.30\%} and \vtwo{26.93\%} regarding clustering effectiveness, in simulated and real-world environments, respectively.

This paper makes the following contributions:

\begin{itemize}
	\item \vtwo{\textbf{A novel type of failure proximity.} We propose the program variable-based failure proximity, which uses run-time program variable information as the failure representer. To the best of our knowledge, this is the first time that program variables serve as proxies for failures in failure indexing.}
	
    \item \vtwo{\textbf{A promising failure indexing approach.} Following the definition of the program variable-based failure proximity, we present a novel failure indexing approach comprising a new fingerprinting function and a tailored distance metric.}

	\item \vtwo{\textbf{A comprehensive evaluation.} We use a diversity of benchmarks and select convincing metrics for the experiments, revealing the competitiveness of the proposed approach.}

\end{itemize}

\section{Background}
\label{sect:back}

\subsection{Parallel Debugging}
\label{subsect:parallel}
Many studies show that fault localization will be more difficult if multiple faults co-exist in a program~\cite{digiuseppe2011influence[6],wang2020[zxh4],keller2017critical[zxh5],xiaobo2018analysis[zxh13],xue2013significant[zxh14]}. A main reason lies in the fault interference~\cite{debroy2009insights[5],wong2016survey[zxh37]}, that is, the phenomenon of the presence of a fault to cause the ineffectiveness of the fault-localization technique to locate another fault~\cite{digiuseppe2011influence[6]}. To tackle this challenge, a natural idea is to localize each fault in an independent environment. Thus, researchers and developers often draw on the idea of parallel debugging, i.e., partitioning failures into groups that target a single fault each (\vtwo{indexing} a failure to its causative fault~\cite{debroy2009insights[5]}). \grz{And parallelization can also promote the debugging efficiency.} \vtwo{As discussed in Section~\ref{sect:intro}, high-quality parallel debugging needs reasonable failure indexing, where the failure proximity, i.e., failure representation and distance measurement, is essential.} Here we introduce these two parts of CC and SD-based failure proximities.

\subsection{Failure Representation}
\label{subsect:represent}
In software testing, test cases are typically in the form of program inputs, while failed test cases are those that produce unexpected outputs. Directly available information of a failed test case only contains two parts, the input (i.e, data fed to the program) and the label (i.e., \emph{failed}). It is quite difficult to index failures with only these two sources, since they are poorly distinguishable.

Actually, during running a test case, diverse run-time information is generated, which could provide failure representation with powerful support to alleviate the mentioned threat. Based on that, a series of fingerprinting functions were proposed. For example:

\begin{itemize}

    \item \textbf{Fingerprinting function of CC-based failure proximity.} The CC-based failure proximity represents a failure as a  numerical vector of program coverage. Specifically, it creates a vector with the length equal to the number of program executable statements, and sets the value of the $i^{\rm{th}}$ element as 1 \vtwo{or the execution frequency} if a failed test case covers the $i^{\rm{th}}$ statement during the execution, and 0 otherwise~\cite{digiuseppe2012software[zxh35], hogerle2014more[zxh36], huang2013empirical[36]}.

    \item \textbf{Fingerprinting function of SD-based failure proximity.} The SD-based failure proximity represents a failure as a  suspiciousness ranking list of program entities. Specifically, given a failed test case and successful test cases, it employs a fault localization technique (generally an SBFL one) to calculate the risk of program entities being faulty, and produces a ranking list in which all entities are descendingly ordered by their suspiciousness~\cite{jones2007debugging[9], liu2006failure[zxh3], gao2019mseer[8]}.
\end{itemize}
\vtwo{Though these two are recognized as the most widespread and promising strategies to date,} the basic source on which they rely is still code coverage, whose limitation has been mentioned in~\cite{mao2014slice[zxh15],golagha2019failure[10],lamraoui2016formula[zxh17]} and will be further revealed in Section~\ref{sect:motivation}. A more effective fingerprinting function in deeper insight remains lacking.

\subsection{Distance Measurement}
\label{subsect:measurement}
Defining a reasonable fingerprinting function is the first step in failure proximity. Once failures are translated into the corresponding proxies, properly measuring the distance among the proxies is of great importance. Designing such a distance metric is not an independent process, since it must match the characteristics of the proxies. For example:

\begin{itemize}

    \item \textbf{Distance metric of CC-based Failure proximity.} The CC-based failure proximity typically utilizes the Euclid distance, since it represents a failure as a program code coverage vector, and the Euclid distance is a simple and cheap way to measure the distance between such numerical vectors.

    \item \textbf{Distance metric of SD-based Failure proximity.} The SD-based failure proximity typically utilizes the Kendall tau distance, the Jaccard distance, \vtwo{and the Euclid distance}, since it represents a failure as a ranking list, which can be suitably handled by these three distance metrics.

\end{itemize}
It can be seen that the design of the distance metric is highly dependent on the fingerprinting function. To put it another way, a work defining a novel type of fingerprinting function should also design a tailored distance metric at the same time.

\section{Motivation}
\label{sect:motivation}

In Section~\ref{sect:intro}, we claim that if the coverage of the failures having distinct root causes is identical, neither of CC-based and SD-based failure proximities can work well. Here we use a motivating example in Listing~\ref{list:example} to illustrate such a scenario.

This program aims to identify and replace certain words from the input string, and then output the modified string and the log message. \syrevised{Specifically, if an input string contains ``\emph{wordNone}'' or ``\emph{wordNtwo}'', these two words will be replaced with ``\emph{*1*}'' and ``\emph{*2*}'', respectively. The log message records the operation of the program.} Given a test suite containing 12 test cases: $t_1$ = ``\emph{speak\ wordNone}'', $t_2$ = ``\emph{wordNone}'', $t_3$ = ``\emph{wordNonecontained}'', $t_4$ = ``\emph{wwwwordNoneeee}'', $t_5$ = ``\emph{has\ wordNtwo}'', $t_6$ = ``\emph{wordNtwo}'', $t_7$ = ``'', $t_8$ = ``\emph{midd*1*le}'', $t_9$ = 
\begin{lstlisting}[language=Java]
	public static String process(String s){
		if(s.contains("*1*") || s.contains("*2*")){
			return "";}
		int sign = 0;
		int sum_1 = 0;
		sum_1 = s.contains("wordNone") ? 1 : 0;
		sign += sum_1;
		s = s.replaceAll("wordNone", "?1?");                                    // Fault1: "?1?" should be "*1*"
		int sum_2 = 0;
		sum_2 = s.contains("wordNtwo") ? 2 : 0;
		sign += sum_2;
		s = s.replaceAll("wordNtwo", "*2*");
		if(sign == 3){
			return "both pattern recognized";}
		String msg = sign == 1 ? "wordNone recognized" : "pass";
		msg = sign > 2 ? "wordNtwo recognized" : msg;                          // Fault2: "> 2" should be "== 2"
		return s + "//" + msg;}
\end{lstlisting}
``\emph{*1*2*}'', $t_{10}$ = ``\emph{a normal sentence}'', $t_{11}$ = ``\emph{wordnonewordNtw}'', and $t_{12}$ = ``\emph{wordNone and wordNtwo}''. Six of them are labeled as $failed$ due to the unexpected outputs: $t_1$, $t_2$, $t_3$ and $t_4$ are triggered by $Fault_1$, $t_5$ and $t_6$ are triggered by $Fault_2$ (we refer to the failed test cases $t_1 \sim t_6$ as $f_1 \sim f_6$, respectively). \syrevised{An ideal failure indexing process should satisfy two goals. The first is to correctly predict the number of clusters (i.e., the number of faults, two). And the second is to properly index each of the failures to their root cause, i.e., delivering two clusters, \{$f_1$, $f_2$, $f_3$, $f_4$\} and \{$f_5$, $f_6$\}.}

\vtwo{We can find that all of these six failed test cases cover the same set of statements, i.e.,  \{$s_1$, $s_2$, $s_4$, $s_5$, $s_6$, $s_7$, $s_8$, $s_9$, $s_{10}$, $s_{11}$, $s_{12}$, $s_{13}$, $s_{15}$, $s_{16}$, $s_{17}$\}. Therefore, the CC-based failure proximity will represent these failures identically thus have no way to distinguish them. Notice that if the gathered coverage information is the same, the SD-based failure proximity will also be trapped. This is because the intuition of SD strategies is that a fault can be triggered in many different ways. The incorporated SBFL techniques cannot handle the scenario that multiple faults are triggered in the same \tbdtwo{path}.}

That is to say, even \vtwo{the most widespread} and the state-of-the-art failure proximities are still not enough to deliver promising outcomes in such a situation, which, \vtwo{unfortunately, is not uncommon in practice.} Thus, developing a better failure proximity is of great significance. To this aim, two questions must be answered: 1) How to define a fingerprinting function to better extract the signature of failures? And 2) How to design a tailored distance metric for measuring the similarity between the proxies for failures?

\section{Approach}
\label{sect:approach}

\begin{figure*}  
	\centering  
	\includegraphics[height=3.9cm,width=16.3cm]{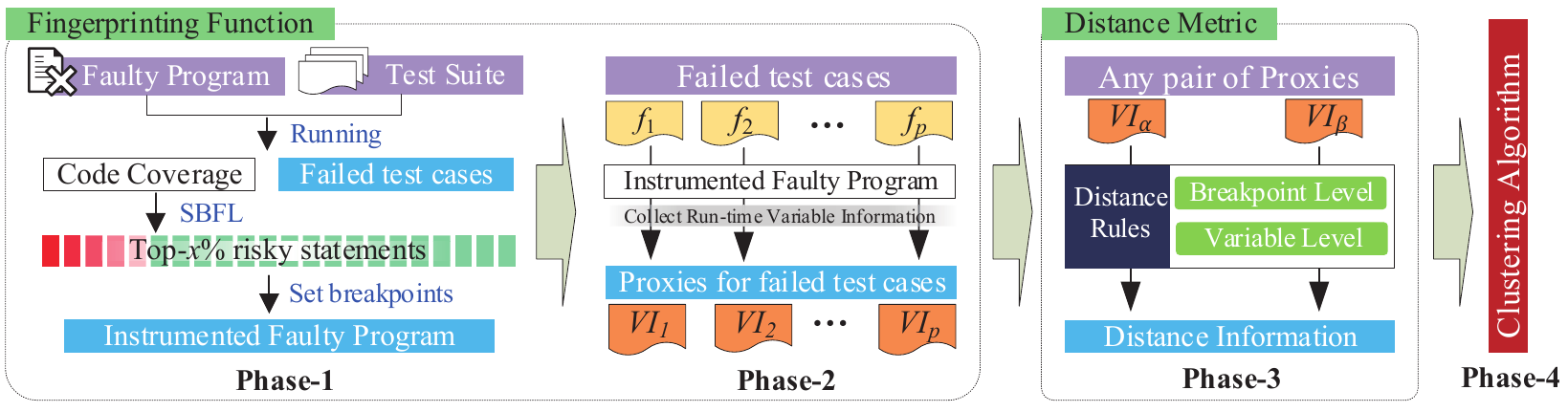} 
	\caption{The overview of the proposed approach} 
	\label{fig:overview}  
\end{figure*}

In this section, we \tbdtwo{propose} \vtwo{the program variable-based failure proximity and} outline the workflow of our novel failure indexing approach in Section~\ref{subsect:overview}, describe the \syrevised{definition of the proposed} fingerprinting function \syrevised{in Section~\ref{subsect:finger}, elaborate on the} tailored distance metric in Section~\ref{subsect:distance}, and introduce the \syrevised{clustering algorithm we employ} in Section~\ref{subsect:clustering}. \syrevised{For smooth understanding, we explain a running example in Section~\ref{subsect:example}.}

\subsection{Overview}
\label{subsect:overview}

\vtwo{The program variable-based failure proximity uses the run-time values of  program variables to represent failures, and measures the similarity between failures based on the characteristic of variable information. Following this description, we propose our novel failure indexing approach.} We depict its overall workflow in Figure~\ref{fig:overview} and summarize the four phases as follows:

\begin{itemize}
    \item \textbf{Phase-1:} Executing the test suite against the faulty program, inputting the code coverage into a spectrum-based fault localization technique to calculate suspiciousness for each program statement, and determining the Top-\textbf{$x$}\% riskiest statements as breakpoints.

    \item \textbf{Phase-2:} For each failed test case $f_i$ ($i$ = 1, 2, ..., $p$), collecting the run-time variable information during its execution at the pre-set breakpoints. The variable information will serve as the proxy for the corresponding failure, denoted as $VI_i$ ($i$ = 1, 2, ..., $p$).

    \item \textbf{Phase-3:} Calculating the distance between each pair of $VIs$ using the proposed distance metric.

    \item \textbf{Phase-4:} Delivering the distance information to the downstream clustering algorithm, enabling all failures to be indexed to their own root cause.
\end{itemize}
\vtwo{Phase-1 and Phase-2 involve the fingerprinting function, Phase-3 involves the distance metric, and Phase-4 is the clustering algorithm. Next we elaborate on the technical details of them.}

\subsection{Fingerprinting Function}
\label{subsect:finger}

Instead of \vtwo{solely} considering code coverage, we dig deeper into available information \syrevised{during the dynamic execution}, and use \syrevised{run-time program variable information, i.e.,} the run-time values of program variables \syrevised{queried at a set of breakpoints during running a failed test case}, to better represent \syrevised{this} failure. \syrevised{It is obvious that such a fingerprinting function involves two factors: the breakpoint determination and the variable information collection.}

\subsubsection{Breakpoint determination (Phase-1)}
\label{subsubsect:breakpoint}

We first employ an SBFL technique to calculate suspiciousness for each program statement, and then determine whether a statement should be set as a breakpoint (the Top-$x$\% riskiest statements are selected). The intuition we adopt this strategy is that statements with higher suspiciousness are more likely to be faulty, and variable information collected at these \syrevised{positions could have stronger capability to revealing the faults, thus can contribute more to representing failures.} \grz{We further investigate the impact of the value of $x$ on the effectiveness of the proposed approach in Section~\ref{subsect:rq1}.}

\subsubsection{Variable information collection (Phase-2)}
\label{subsubsect:varcollection}

\syrevised{In the context of the proposed approach, the proxy for a failure is defined as a two-dimension dictionary. Specifically, the $i^{\rm{th}}$ failed test case $f_i$ in the test suite can be represented as \vtwo{$VI_i$ = \{$bp_1$: $V^i_1$, $bp_2$: $V^i_2$, ..., $bp_j$: $V^i_j$,..., $bp_q$: $V^i_q$\}}, where $V^i_j$} denotes the variable information queried at the $j^{\rm{th}}$ breakpoints (denoted as $bp_j$) during the execution of $f_i$, $q$ is the total number of pre-set breakpoints. $V^i_j$ is also a dictionary, which contains the name (dictionary's key) and value (dictionary's value) of the queried variables.

\tbd{When a program stops at a statement, only the variable information at the position before the execution of that statement can be collected. Thus, to make sure that variable information is collected regarding the positions determined by Phase-1, when the program stops at each breakpoint, we continue executing a further step and then carry out the collection operation.} \syrevised{In addition, if a statement is covered more than once, the variable information is only collected in the last execution to keep it up-to-date.}

\subsection{Distance Metric}
\label{subsect:distance}

\syrevised{In failure indexing, a basic concept is that \emph{failures triggered by the same fault should be as similar as possible, and vice versa}. Thus, in the context of the proposed approach, the intuition of designing a distance metric \vtwo{can be concretized as} \emph{the run-time variable information of the failures caused by the same fault should be as similar as possible, and vice versa}.} For making the proposed approach better competent to such a mission, its distance metric is designed with two levels: the breakpoint level and the variable level. \syrevised{Given a pair of failures that requires measuring the distance, $f_\alpha$ and $f_\beta$, the breakpoint level divides all of the breakpoints into three categories, i.e, those covered by both $f_\alpha$ and $f_\beta$, those covered by only one of $f_\alpha$ and $f_\beta$, and those covered by neither of them.} Then, the variable level \syrevised{further} compares the variable information at \syrevised{the breakpoints fallen into the first category}. The detailed descriptions of the two levels are as follows.

\subsubsection{Breakpoint level (Phase-3)}
\label{subsubsect:bplevel}

\syrevised{Given a pair of failed test cases, $f_\alpha$ and $f_\beta$, the breakpoint level calculates the distance between them using Formula~\ref{equ:dis_ab},}
\begin{equation}
\label{equ:dis_ab}
Distance_{f_\alpha,\ f_\beta}=\frac{\sum_{j}^{q}Distance_{bp_j}}{\sum_{j}^{q}BPCount_j}
\end{equation}
where $q$ is the number of pre-set breakpoints, $Distance_{bp_j}$ is the distance between $f_\alpha$ and $f_\beta$ at the $j^{\rm{th}}$ breakpoint $bp_j$, and $BPCount_j$ is a binary constant for getting the mean of distances\footnote{There could be an uncommon situation that the value of $\sum_{j}^{q}BPCount_j$ is zero, which means that two failures do not cover any breakpoint. Setting more breakpoints can handle it, and we investigate this question in Section~\ref{subsect:rq1}.}.

\syrevised{For $Distance_{bp_j}$, we calculate its value according to both Formula~\ref{equ:dis_bpj} and Formula~\ref{equ:ej_ab}.}

\begin{equation}
\label{equ:dis_bpj}
Distance_{bp_j}=\left
\{\begin{aligned}
&Distance_{var}^j & \quad e_j^\alpha +e_j^\beta=2 \\
& \qquad 1 & \quad e_j^\alpha+e_j^\beta=1 \\
& \qquad 0 & \quad e_j^\alpha+e_j^\beta=0
\end{aligned}
\right.
\end{equation}

\begin{equation}
\label{equ:ej_ab}
e_j^{\alpha/\beta}= \begin{cases}1 & \quad \text {if } f_\alpha/f_\beta \text { covers } bp_j \\ 0 & \quad \text {if } f_\alpha/f_\beta \text { not covers } bp_j\end{cases}
\end{equation}

\syrevised{And for $BPCount_j$, we calculate its value according to both Formula~\ref{equ:bpcount} and Formula~\ref{equ:ej_ab}.}

\begin{equation}
\label{equ:bpcount}
BPCount_j=\left\{\begin{array}{cc}
1 & \quad e_j^\alpha+e_j^\beta=2 \\
1 & \quad e_j^\alpha+e_j^\beta=1 \\
0 & \quad e_j^\alpha+e_j^\beta=0
\end{array}\right.
\end{equation}

\syrevised{It can be seen that all breakpoints are divided into three categories, and as a consequence of which, the calculation of $Distance_{f_\alpha,\ f_\beta}$ also involves three scenarios:}

\begin{itemize}
\item \textbf{For those breakpoints covered by both of the failures \syrevised{($e_j^\alpha+e_j^\beta=2$)}}. $Distance_{bp_j}$ is set to $Distance_{var}^j$, which will be further calculated based on the variable information, and such a process is discussed at the variable level \syrevised{in Section~\ref{subsubsect:varlevel}}. And $BPCount_j$ is set to 1.

\item \textbf{For those breakpoints covered by only one of the failures \syrevised{($e_j^\alpha+e_j^\beta=1$)}}. It means that two failures have \syrevised{distinct} execution paths at those breakpoints, their variable information at those breakpoints should also be regarded as distinct. \syrevised{Thus,} $Distance_{bp_j}$ is \syrevised{directly} set to the maximum value, 1 (all the values of $Distance_{bp_j}$ will be normalized to the interval of [0, 1]). And $BPCount_j$ is set to 1.

\item \textbf{For those breakpoints covered by neither of the failures \syrevised{($e_j^\alpha+e_j^\beta=0$)}}. We think \syrevised{it is difficult for them to make any contribution to failure indexing}. Therefore, $Distance_{bp_j}$ is \syrevised{directly} set to 0, and $BPCount_j$ is also set to 0, to make sure they have no impact on the outcome of $Distance_{f_\alpha,\ f_\beta}$.
\end{itemize}

\subsubsection{Variable level (Phase-3)}
\label{subsubsect:varlevel}

\syrevised{As defined in Formula~\ref{equ:dis_bpj}, if failed test cases $f_\alpha$ and $f_\beta$ both cover $bp_j$, $Distance_{bp_j}$ will be set to $Distance_{var}^j$, which is with in the scope of the variable level. We calculate the value of $Distance_{var}^j$ using Formula~\ref{equ:dis_var},}

\begin{equation}
\label{equ:dis_var}
Distance_{var}^j= \begin{cases}\frac{\sum_{z}^{\vert\hat{V}_{j}^{\alpha} \cup \hat{V}_{j}^{\beta}\vert} dis_z}{\vert\hat{V}_{j}^{\alpha} \cup \hat{V}_{j}^{\beta}\vert} & \text { if }\vert\hat{V}_{j}^{\alpha} \cup \hat{V}_{j}^{\beta}\vert>0 \\ \qquad \qquad 1 & \text { if }\vert\hat{V}_{j}^{\alpha} \cup \hat{V}_{j}^{\beta}\vert==0\end{cases}
\end{equation}
where $\hat{V}_{j}^{\alpha/\beta}$ is the set of all variables' names collected at $bp_j$ while executing $f_{\alpha}$ or $f_{\beta}$. $\hat{V}_{j}^{\alpha} \cup \hat{V}_{j}^{\beta}$ stands for the \textbf{union} of all variables' names collected at $bp_j$ while executing $f_{\alpha}$ and $f_{\beta}$, and $\vert\hat{V}_{j}^{\alpha} \cup \hat{V}_{j}^{\beta}\vert$ is the scale of this union. If the value of $\left|\hat{V}_{j}^{a} \cup \hat{V}_{j}^{b}\right|$ equals 0, meaning that no variable is collected at $bp_j$ while executing the two failures, we simply set $Distance_{var}^j$ to the maximum value (i.e., 1) to handle this situation. Otherwise, We use a $variable$-$to$-$variable$ tactic to measure the similarity between $f_{\alpha}$ and $f_{\beta}$ at $bp_j$. \syrevised{Specifically,} \syrevised{each of the variables in $\hat{V}_{j}^{\alpha} \cup \hat{V}_{j}^{\beta}$ will be compared with its counterpart (i.e., the same variable collected during executing the other failure).}

\syrevised{Such a $variable$-$to$-$variable$ tactic is implemented by $dis_z$ in Formula~\ref{equ:dis_var}}, which is the distance between the values of the $z^{\rm{th}}$ variable in the execution of $f_{\alpha}$ and $f_{\beta}$, as defined in Formula~\ref{equ:dis_z}, Formula~\ref{equ:cz_ab}, and Formula~\ref{equ:nz_ab} (For convenience, we denote the $z^{\rm{th}}$ variable as $var_z$, and \vtwo{denote the values of $var_z$ during the execution of $f_{\alpha}$ and $f_{\beta}$ as $val_z^{\alpha}$ and $val_z^{\beta}$, respectively).}

\begin{equation}
\label{equ:dis_z}
\operatorname{dis}_{z}=\left\{\begin{array}{lr}
\operatorname{Jacc}\left(val_{z}^{\alpha},\ val_{z}^{\beta}\right) & c_{z}^{\alpha}+c_{z}^{\beta}=2\ \textbf{and}\ n_{z}^{\alpha}+n_{z}^{\beta}=0 \\

\qquad \qquad 1 & c_{z}^{\alpha}+c_{z}^{\beta}=2\ \textbf{and}\ n_{z}^{\alpha}+n_{z}^{\beta}=1 \\

\qquad \qquad 0 & c_{z}^{\alpha}+c_{z}^{\beta}=2\ \textbf{and}\ n_{z}^{\alpha}+n_{z}^{\beta}=2 \\

\qquad \qquad 1 & c_{z}^{\alpha}+c_{z}^{\beta}=1
\end{array}\right.
\end{equation}

\begin{equation}
\label{equ:cz_ab}
c_{z}^{\alpha/\beta}= \begin{cases}1 & \text { if } f_{\alpha/\beta} \text { collects } var_{z} \\ 0 & \text { if } f_{\alpha/\beta} \text { not collects } var_{z}\end{cases}
\end{equation}

\begin{equation}
\label{equ:nz_ab}
n_{z}^{\alpha/\beta}= \begin{cases}0 & \text { if } val_{z}^{\alpha} \text{ / } val_{z}^{\beta}  \text { is not null } \\ 1 & \text { if } val_{z}^{\alpha} \text{ / } val_{z}^{\beta} \text { is null }\end{cases}
\end{equation}

\syrevised{It can be seen that all of the variables are divided into four categories, and as a consequence of which, the calculation of $dis_z$ also involves four scenarios:}

\begin{itemize}
\item \textbf{If a variable is collected by both of the failures, and neither of their values are null} \syrevised{($c_{z}^{\alpha}$+$c_{z}^{\beta}$=2 and $n_{z}^{\alpha}$+$n_{z}^{\beta}$=0)}, we use the Jaccard distance, which is defined in Formula~\ref{equ:jacc}, to calculate $dis_z$.

\item \textbf{If a variable is collected by both of the failures, and one of the values is null}, \syrevised{($c_{z}^{\alpha}$+$c_{z}^{\beta}$=2 and $n_{z}^{\alpha}$+$n_{z}^{\beta}$=1)}, we assign the maximum value (i.e., 1) to $dis_z$, since the two values are uncomparable, which shows the divergence between the two failures when it comes to this variable at $bp_j$.

\item \textbf{If a variable is collected by both of the failures, and both of the two values are null} \syrevised{($c_{z}^{\alpha}$+$c_{z}^{\beta}$=2 and $n_{z}^{\alpha}$+$n_{z}^{\beta}$=2)}, we assign the minimum value (i.e., 0) to $dis_z$, since such a condition can be considered as $val_{z}^{\alpha}$ and $val_{z}^{\beta}$ having no difference.

\item \textbf{If a variable is collected by only one of the failures} \syrevised{($c_{z}^{\alpha}$+$c_{z}^{\beta}$=1)}, we assign the maximum value (i.e., 1) to $dis_z$, because this condition indicates that the two failures may have different dataflows. 
\end{itemize}

\syrevised{As mentioned above, if $var_z$ falls into the first category, the value of $dis_z$ is calculated by \vtwo{the Jaccard distance using Formula~\ref{equ:jacc}. \grz{This is because in our experiments, we find that regarding a variable's value as a string regardless of its original type is beneficial for distance measurement.}}}
\begin{equation}
\label{equ:jacc}
\operatorname{Jacc}\left(val_{z}^{\alpha},\ val_{z}^{\beta}\right)=\text{norm}\left(1\ \text{-}\ \frac{\vert val_{z}^{\alpha} \cap val_{z}^{\beta}\vert}{\vert val_{z}^{\alpha} \cup val_{z}^{\beta}\vert}\right)
\end{equation}

The function $norm$ is used to achieve the 0-1 normalization, which is \syrevised{employed to make the same scale for the distances in the four scenarios, and is defined in Formula~\ref{equ:norm}.}

\begin{equation}
\label{equ:norm}
\operatorname{norm}(J_z)=\frac{{J_z}-\min(J)}{\max(J) -\min(J) }
\end{equation}
Where $J_z$ is the value of $Jacc\left(val_{z}^{\alpha},\ val_{z}^{\beta}\right)$ without the normalization. And $max$($J$) and $min$($J$) are the maximum and the minimum values among all $J_z$, respectively.

\begin{figure}  
	\centering  
	\includegraphics[height=3.3cm,width=6.6cm]{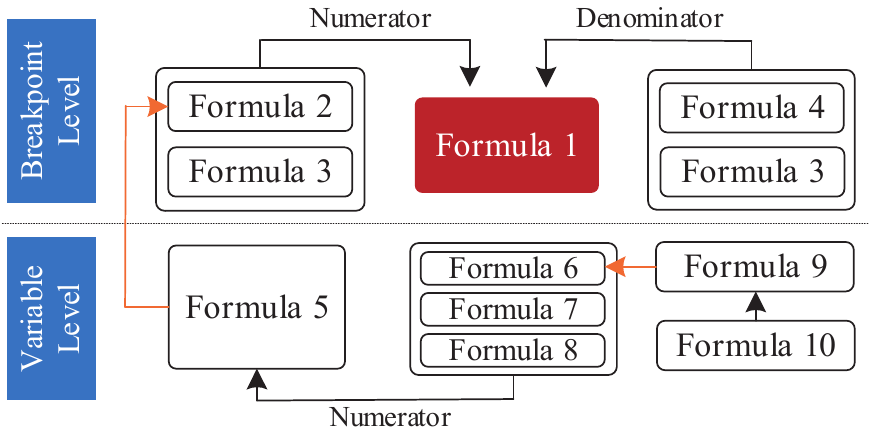} 
	\caption{The workflow of the distance metric} 
	\label{fig:distance}  
\end{figure}

\vtwo{Totally speaking, }\syrevised{Figure~\ref{fig:distance} depicts the workflow of the distance metric, clarifying the relationship among all of the mentioned formulas: Formula~\ref{equ:dis_ab} is the core of the distance metric, while Formula~\ref{equ:dis_bpj} and Formula~\ref{equ:ej_ab} are for its numerator, and Formula~\ref{equ:bpcount} and Formula~\ref{equ:ej_ab} are for its denominator. Formula~\ref{equ:dis_var} is called by Formula~\ref{equ:dis_bpj}, which can be supported by Formula~\ref{equ:dis_z}, Formula~\ref{equ:cz_ab}, and Formula~\ref{equ:nz_ab}. And Formula~\ref{equ:dis_z} is determined by Formula~\ref{equ:jacc} and Formula~\ref{equ:norm}.}

\subsection{Clustering Algorithm}
\label{subsect:clustering}
We \syrevised{employ the clustering component in} MSeer~\cite{gao2019mseer[8]}, the most up-to-date failure indexing technique, \syrevised{to complete the clustering stage of the proposed approach (i.e., Phase-4 in Figure~\ref{fig:overview})}. \syrevised{This algorithm involves the faults number estimation and the clustering. The former is responsible for the prediction of the number of clusters, and the latter identifies the due-to relationship between the observed failures and the underlying faults. Next we give a concise description and more details can be found in~\cite{gao2019mseer[8]}.}

\subsubsection{The faults number estimation (Phase-4)}
\label{subsubsect:estimation}
\syrevised{It is well-recognized that one of the trickiest challenges in clustering lies in the estimation of the number of clusters~\cite{tibshirani2001estimating[zxh18],kingrani2018estimating[zxh19],fu2020estimating[zxh20]}. Putting it into the context of failure indexing, we can claim that predicting the number of faults given a set of failures is of great importance. \vtwo{The adopted algorithm} presents a novel mountain method-based technique inspired by previous works~\cite{yager1994approximate[14], chiu1994fuzzy[15]}, to carry out the faults number estimation and the assignment of initial medoids to clusters simultaneously. Specifically, \vtwo{it} first calculates a potential value for each data point (i.e., a failure) according to the density of its surrounding, such a potential value is used to measure the possibility of a failure being set as a medoid. And then, 1) Choosing the failure with the highest potential value as a medoid. 2) Updating the potential values of all failures in accordance with their distance from the newest medoid. 3) Repeating these two processes iteratively, until the maximum potential value falls within a certain threshold.}

\subsubsection{The clustering (Phase-4)}
\label{subsubsect:clustering}
\syrevised{Once the number of clusters and the initial medoids are determined, all failures are ready to be clustered. \vtwo{The adopted algorithm} utilizes K-medoids, a widely-used clustering technique, to complete this process. The K-medoids technique sets \vtwo{actual} (rather than virtual) data points as medoids thus can be more applicable to the proposed approach, because the mean of variable information is difficult to define. The K-medoids technique has also shown to be very robust to the existence of noise or outliers~\cite{Kaufman[38]}.}

\subsection{Running Example}
\label{subsect:example}

\vtwo{We recall the example in Listing~\ref{list:example} to exemplify the workflow of the proposed approach, highlighting its capability to handling the scenario where CC and SD strategies are trapped.}

\textbf{For Phase-1} (Breakpoint determination). Employing an SBFL technique (e.g., DStar~\cite{wong2013dstar[3]}) to calculate suspiciousness for each statement, and sorting all statements in descending order of suspiciousness: \{$s_{15}, s_{16}, s_{17}, s_{4}, s_{5}, ...\}$. If several statements share the same value of suspiciousness, we adopt the widely-used solution by ranking them in the ascending order of line numbers~\cite{pearson2016evaluation[28], sun2016properties[29], xu2011ties[30], gao2019mseer[8]}. The proposed approach will determine several riskiest statements as breakpoints (the determination threshold will be investigated in Section~\ref{subsect:rq1}, here we take Top-10\% as an example). Thus, $s_{15}$ and $s_{16}$ are selected as breakpoints.

\textbf{For Phase-2} (Variable information collection). Collecting the run-time variable information at \vtwo{$s_{15}$ ($bp_1$) and $s_{16}$ ($bp_2$)} during the execution of the six failed test cases. Revisiting the definition in Section~\ref{subsubsect:varcollection}, we can represent failed test cases $f_1 \sim f_6$ as the run-time variable information $VI_1 \sim VI_6$, respectively. Taking $f_1$ (triggered by $Fault_1$) and $f_5$ (triggered by $Fault_2$) as an example, the proxies for $f_1$ and $f_5$ are $VI_1$ = \{$bp_1$: $V^1_1$,  $bp_2$: $V^1_2$\} and $VI_5$ = \{$bp_1$: $V^5_1$,  $bp_2$: $V^5_2$\}, respectively. $V^1_1$, $V^1_2$, $V^5_1$, and $V^5_2$ are given as follows\footnote{\vtwo{During executing $f_1$ and $f_5$, the variable information collected at the two breakpoints is identical, since the scale and complexity of the used toy program are very low.}}.

\begin{center}\small
	\begin{tcolorbox}[colback=gray!15,
		colframe=black,
		width=8.2cm,
		arc=1mm, auto outer arc,
		boxrule=0.5pt,
		]
		{
			\hangafter=1
			\setlength{\hangindent}{2cm}
			$V^1_1$ = $V^1_2$ = \{$s$: ``speak ?1?'', $sign$: ``1'', $sum\_1$: ``1'', $sum\_2$: ``0'', \text{\qquad\qquad\ \ \ \ }$msg$: ``wordNone recognized''\}\\
			$V^5_1$ = $V^5_2$ = \{$s$: ``has *2*'', $sign$: ``2'', $sum\_1$: ``0'', $sum\_2$: ``2'', \text{\qquad\qquad\ \ \ \ \ }$msg$: ``pass''\}}
	\end{tcolorbox}
\end{center}

\vtwo{The evident distinction between the variable information of $f_1$ and that of $f_5$ exhibits that these two failures have different dataflows, despite the fact that they have the same coverage.}

\textbf{For Phase-3} (Distance measurement). A failed test case has been converted to the variable information \vtwo{in} Phase-1 and Phase-2. Therefore, the distance measurement between two failures equals that between two sets of variable information. The distances between each pair of failures are calculated and given in Table~\ref{tab:distance}. Revisiting the mapping relationship between the two faults and the six failures, i.e., $Fault_1: \{f_1, f_2, f_3, f_4\}$ and $Fault_2: \{f_5, f_6\}$. It can be seen that the failed test cases triggered by the same fault are highly similar, while on the contrary, the failed test cases triggered by different faults show low similarity to each other.

\begin{table}\small
	\caption{\label{tab:distance} Distance information of the running example} 
	\centering
	\begin{tabular}{|c|c|c|c|c|c|c|} 
		\hline
		
		& \bm{$f_{1}$}  & \bm{$f_{2}$}  & \bm{$f_{3}$}  & \bm{$f_{4}$}  & \bm{$f_{5}$}  & \bm{$f_{6}$}                      \\ 
		\hline
		\bm{$f_{1}$} & 0   & \cellcolor{gray!20}0.2 & \cellcolor{gray!20}0.2 & \cellcolor{gray!20}0.2 & \cellcolor{gray!50}0.8 & \cellcolor{gray!50}1                       \\ 
		\hline
		\bm{$f_{2}$} & \cellcolor{gray!20}0.2 & 0   & \cellcolor{gray!20}0.2 & \cellcolor{gray!20}0.2 & \cellcolor{gray!50}1   & \cellcolor{gray!50}1                       \\ 
		\hline
		\bm{$f_{3}$} & \cellcolor{gray!20}0.2 & \cellcolor{gray!20}0.2 & 0   & \cellcolor{gray!20}0.2 & \cellcolor{gray!50}0.8 & \cellcolor{gray!50}1                       \\ 
		\hline
		\bm{$f_{4}$} & \cellcolor{gray!20}0.2 & \cellcolor{gray!20}0.2 & \cellcolor{gray!20}0.2 & 0   & \cellcolor{gray!50}1   & \cellcolor{gray!50}1                       \\ 
		\hline
		\bm{$f_{5}$} & \cellcolor{gray!50}0.8 & \cellcolor{gray!50}1   & \cellcolor{gray!50}0.8 & \cellcolor{gray!50}1   & 0   & \cellcolor{gray!20}0.2                     \\ 
		\hline
		\bm{$f_{6}$} & \cellcolor{gray!50}1   & \cellcolor{gray!50}1   & \cellcolor{gray!50}1   & \cellcolor{gray!50}1   & \cellcolor{gray!20}0.2 & 0                       \\
		\hline
	\end{tabular}
\end{table}

\textbf{Phase-4} (Clustering). Running the clustering algorithm, two clusters can be obtained: $\{f_1, f_2, f_3, f_4\}$ and $\{f_5, f_6\}$. This result indicates that our approach can properly divide failures according to their root cause, delivering a promising failure indexing process.

\vtwo{From this example we can observe that the mechanism of the proposed approach lies in the dataflow. Specifically, when the coverage of the failures having distinct root causes is identical, the internal dataflow (the run-time values of program variables in our context) could play a role of distinguisher. It should be pointed out that although dynamic slices can reflect the dataflow to an extent thus can preliminarily divide those failures, we are interested in whether there is a finer-grained and more precise representer that can be competent to such a mission.}

\section{EXPERIMENTAL SETUP}
\label{sect:setup}

In this section, we introduce the experimental setup of this study, including \vtwo{research questions, }parameter setting, datasets, metrics, and environments.

\subsection{Research Questions}
\label{subsect:rq}

\begin{itemize}
	\item \textbf{\vtwo{RQ1: The value of the hyperparameter.}} In the fingerprinting function, the proposed approach determines the Top-$x$\% riskiest statements as breakpoints. We investigate how the value of $x$ impact the effectiveness of our approach.
	
	\item \textbf{\vtwo{RQ2: Impact analyses of components.}} In the distance metric, the proposed approach measures the similarity between two sets of variable information through the breakpoint level and the variable level. How does each of the two components impact the effectiveness of our approach?
	
	\item \textbf{\vtwo{RQ3: Competitiveness of the proposed approach.}} How does our approach perform compared with the current most prevalent and promising failure indexing techniques?
\end{itemize}

\subsection{Parameter Setting}
\label{subsect:parameter}

\vtwo{As mentioned in Section~\ref{subsubsect:breakpoint}, we need to first determine an SBFL technique to calculate suspiciousness for each program statement. In the experiment, we choose DStar\footnote{Considering the preference for DStar in many other studies (such as~\cite{pearson2017evaluating[42], arrieta2018spectrum[43], widyasari2022real[45]}), we set the value of * in DStar to 2, the most thoroughly-explored value, in our experiments.}, one of the best SBFL techniques~\cite{wong2013dstar[3]}. Notice that such a choice is not hard-coded but can be configurable, our approach can adapt to any other fault localization techniques that are able to deliver a suspiciousness ranking of program entities at a statement granularity.}

\subsection{Datasets}
\label{subsect:datasets}

\begin{table}[]\small
	\centering
	\setlength{\abovecaptionskip}{0pt}
	\caption{\label{tab:benchmark} \vtwo{Benchmarks}}
	\begin{tabular}{lcccl}
		\hline
		\textbf{Language} & \textbf{Project} & \textbf{Version} & \textbf{kLOC}   & \textbf{Functionality}  \\ \hline
		\multirow{4}{*}{C}        &   flex             & 2.5.3            & 14.5            & Parser generator \\
		        &   grep             & 2.4              & 13.5           & Text matcher \\
		        &   gzip             & 1.2.2            & 7.3             & File archiver \\
		        &   sed              & 3.02             & 10.2           & Stream editor \\
		        
		\rule{0pt}{9pt}\multirow{5}{*}{Java}   &   Chart           & 2.0.0            & 96.3          & Chart library   \\
		   &   Closure        & 2.0.0            & 90.2          & Closure compiler  \\
		   &   Lang            & 2.0.0            & 22.1            & Apache commons-lang  \\
		   &   Math            & 2.0.0            & 85.5           & Apache commons-math \\
		   &   Time            & 2.0.0            & 28.4           & Date and time library \\
		\hline	
	\end{tabular}
\end{table}

\subsubsection{Simulated Scenarios}
\label{subsubsect:simulated}

SIR (Software-artifact Infrastructure Repository) is a classical platform for experiments in software testing and debugging~\cite{do2005supporting[12]}. We download four C projects from SIR: $flex$, $grep$, $gzip$, and $sed$, and then based on which create 1-bug, 2-bug, 3-bug, 4-bug, and 5-bug faulty versions (i.e., faulty programs containing one, two, three, four, and five bugs, respectively) by employing mutation strategies~\cite{papadakis2019mutation[22]}, in light of the fact that previous research has confirmed that mutation-based faults can provide credible results for experiments in software testing and debugging~\cite{andrews2005mutation[16], andrews2006using[17], do2006use[18], just2014mutants[19], liu2006statistical[20], pradel2018deepbugs[21]}. To create an $r$-bug faulty version ($r$ = 1, 2, 3, 4, 5), we inject 1, 2, 3, 4, and 5 mutant(s) into the clean program, respectively. We employ an existing tool with \vtwo{11} ``fork'' and \vtwo{22} ``star'' on GitHub to perform mutation~\cite{mutationtool[23]}. It defines 67 types of points that can be mutated, and provides several mutation operators for each one. For example, replacing operators such as addition, subtraction, multiplication, division, etc. with each other~\cite{jeffrey2008fault[24]}, and reversing an $if$-$else$ predicate, deleting an $else$ statement, modifying a decision condition~\cite{xuan2016nopol[25]}, and so on. \vtwo{The description of the four projects is given in Table~\ref{tab:benchmark}. In total, we create 600 SIR faulty versions.}
 
\subsubsection{Real-world Scenarios}
\label{subsubsect:realworld}

Defects4J is one of the most popular benchmarks in the current field of software testing and debugging, due to its realism and ease-to-use~\cite{just2014defects4j[13]}. We download five Java projects from Defects4J: $Chart$, $Closure$, $Lang$, $Math$, and $Time$, and then based on which search for 1-bug, 2-bug, 3-bug, 4-bug, and 5-bug faulty versions. Notice that Defects4J is typically for single-fault scenarios, namely, no matter how many bugs are contained in a faulty program, the provided test suite is only sufficient to reveal one of them. Such a characteristic hinders its use in failure indexing. To adapt Defects4J to multi-fault scenarios, An et al. presented a search strategy to enhance the test suite. Specifically, they transplant the fault-revealing test case(s) of another faulty version (or other faulty versions) to a basic faulty version, that is, allowing a more robust test suite to find more faults in the original program (i.e., the basic faulty version)~\cite{an2021searching[26]}. \vtwo{The description of the five projects is given in Table~\ref{tab:benchmark}. In total, we get 100 Defects4J faulty versions.}

\subsection{Metrics}
\label{subsect:metrics}
Generally speaking, the capability of a failure indexing approach can be measured from two aspects. One is the faults number estimation, namely, to what extent the number of faults can be correctly predicted given a series of observed failures. And the other is the clustering effectiveness, namely, to what extent these failures can be indexed to their root cause.

\subsubsection{Faults number estimation}
\label{subsubsect:number}
For an $r$-bug faulty version, we utilize a failure indexing approach $T$ to estimate the number of faults $r$. \vtwo{If the estimated number of faults $k$ is equal to $r$, we mark this faulty version as $equal$, and use $V^T_{equal}$ to denote the number of faulty versions that fall into the $equal$ category when using $T$. Obviously, a larger value of $V^T_{equal}$ indicates a stronger capability to representing failures of $T$.}

\subsubsection{Clustering effectiveness}
\label{subsubsect:effectiveness}

\begin{table}[]\small
	\centering
	\setlength{\abovecaptionskip}{0pt}
	\caption{\label{tab:metrics} Scenarios in two types of metrics}
	\begin{tabular}{cc|ll}
		\hline
		\multirow{2}{*}{\textbf{Metric}} & \multirow{2}{*}{\textbf{Notation}} & \multicolumn{2}{c}{\textbf{Results of failure indexing}}  \\ \cline{3-4}  & &
		In generated cluster & In oracle cluster \\ \hline
        \multirow{4}{*}{\textbf{FMI and JC}}& SS                       & Same                            & Same                         \\
		& SD                      & Same                            & Difference              \\
		& DS                      & Difference                  & Same                         \\
		& DD                      & Difference                 & Difference       \\ 
		\rule{0pt}{9pt}\multirow{4}{*}{\textbf{PR and RR}}& TP                       & Positive                     & Positive                         \\
		& FP                       & Positive                     & Negative               \\
		& TN                      & Negative                   & Negative                         \\
		& FN                      &Negative                    & Positive        \\ \hline         
	\end{tabular}
\end{table}

We employ the Fowlkes and Mallows Index (FMI), the Jaccard Coefficient (JC), the Precision Rate (PR), and the Recall Rate (RR), to measure the effectiveness of a clustering process. These four metrics are classic and easily-available, and they have also been adopted in a collection of prior research~\cite{wu2009adapting[zxh39],xiejy2017newcriteria[zxh40], zhang2022smart[37]}.

Among them, FMI and JC compare the indexing consistency of each pair of failed test cases in the generated cluster with that in the oracle cluster~\cite{flynt2018exploration[zxh41]}, as shown in Formula~\ref{equ:fmi} and Formula~\ref{equ:jc}. The four possible scenarios in the comparison can be found in Table~\ref{tab:metrics}.

\begin{equation}
\label{equ:fmi}
FMI = \sqrt{\frac{X_{SS}}{X_{SS} + X_{SD}} \times \frac{X_{SS}}{X_{SS} + X_{DS}}}
\end{equation}

\begin{equation}
\label{equ:jc}
JC = \frac{X_{SS}}{X_{SS} + X_{SD} + X_{DS}}
\end{equation}
\sy{Where $X_{SS}$ denotes the number of pairs of ``SS'', and so forth.}

PR and RR compare the classification result of failed test cases in the generated cluster with that in the oracle cluster, as shown in Formula~\ref{equ:pr} and Formula~\ref{equ:rr}. The four possible scenarios in the comparison can be found in Table~\ref{tab:metrics}.

\begin{equation}
\label{equ:pr}
PR = \frac{X_{TP}}{X_{TP} + X_{FP}}
\end{equation}

\begin{equation}
\label{equ:rr}
RR = \frac{X_{TP}}{X_{TP} + X_{FN}}
\end{equation}
\sy{Where $X_{TP}$ denotes the number of failures of ``TP'', and so forth.}

\syrevised{We deliver only \vtwo{the faulty versions whose number of faults is correctly predicted (hereafter, simply referred to as ``$k$ == $r$'' faulty versions)} to the following clustering phase. The reason behind such a strategy is that, if the predicted number of faults $k$ is not equal to $r$, it is hard to compare the $k$ generated clusters with the $r$ oracle groups. As a consequence of which, the measurement of clustering effectiveness can be difficult. \vtwo{We use Formula~\ref{equ:sum} to calculate the sum of the metric values on ``$k$ == $r$'' faulty versions,}}
\begin{equation}
	\label{equ:sum}
	{S\_M}_{\ M}^{\ T}=\sum_{i}^{V_{equal}^{T}} M_{i}
\end{equation}
where $V^T_{equal}$ is the number of ``$k$ == $r$'' faulty versions when using $T$. $M_i$ is the value of the clustering metric $M$ ($M$ takes FMI, JC, PR, or RR) on the $i^{\rm{th}}$ ``$k$ == $r$'' faulty version.

\sy{Notice that ``$k$ == $r$'' is just an ideal scenario (not necessary) for the proposed approach. Even if $k$ $\neq$ $r$, our approach can also work, as introduced in Section~\ref{sect:intro} (the part of ``over-division'' and ``under-division'').}

\subsection{Environments}
\label{subsect:environments}
We generate faulty versions, collect program coverage and run-time variable information on Ubuntu 16.04.1 LTS with GCC 5.4.0 and JDB 1.8. The distance measurement and clustering processes run on a server equipped with 96 Intel Xeon(R) Gold 5218 CPU cores with 2.30GHz and 160 GB of memory.

\section{RESULT AND ANALYSIS}
\label{sect:result}

\subsection{RQ1: The value of the hyperparameter}
\label{subsect:rq1}
\vtwo{We investigate the effectiveness of the Proposed Approach (PA for short) for the value of $x$ taking 5, 10, 15, and 20 (i.e., determining the Top-5\%, Top-10\%, Top-15\%, and Top-20\% riskiest statements as breakpoints, respectively) \sy{on SIR faulty versions}. The results are given in Table~\ref{tab:rq1} and Figure~\ref{fig:rq1}.}

\subsubsection{The capability to estimating the number of faults in different values of $x$.}
\label{subsubsect:rq1_1}

\vtwo{A promising failure indexing approach should make the number of faults it predicts $k$ equal to the real number of faults $r$. The values of $V^{T}_{equal}$, i.e., on how many faulty versions can ``$k$ == $r$'' be obtained using the proposed approach equipped with  different values of $x$ ($T$ takes PA$_{5\%}$, PA$_{10\%}$, PA$_{15\%}$, and PA$_{20\%}$), are given in Table~\ref{tab:rq1} and Figure~\ref{fig:rq1}. It can be seen that when the breakpoint determination threshold is set to 10\%, the proposed approach can correctly estimate the number of faults on 98 faulty versions, while when such a threshold is set to 5\%, 15\%, and 20\%, the numbers of ``$k$ == $r$'' faulty versions are 75, 83, and 84, respectively. }

\subsubsection{The capability to clustering in different values of $x$.}
\label{subsubsect:rq1_2}
\begin{figure}  
	\includegraphics[height=3.5cm,width=7.5cm]{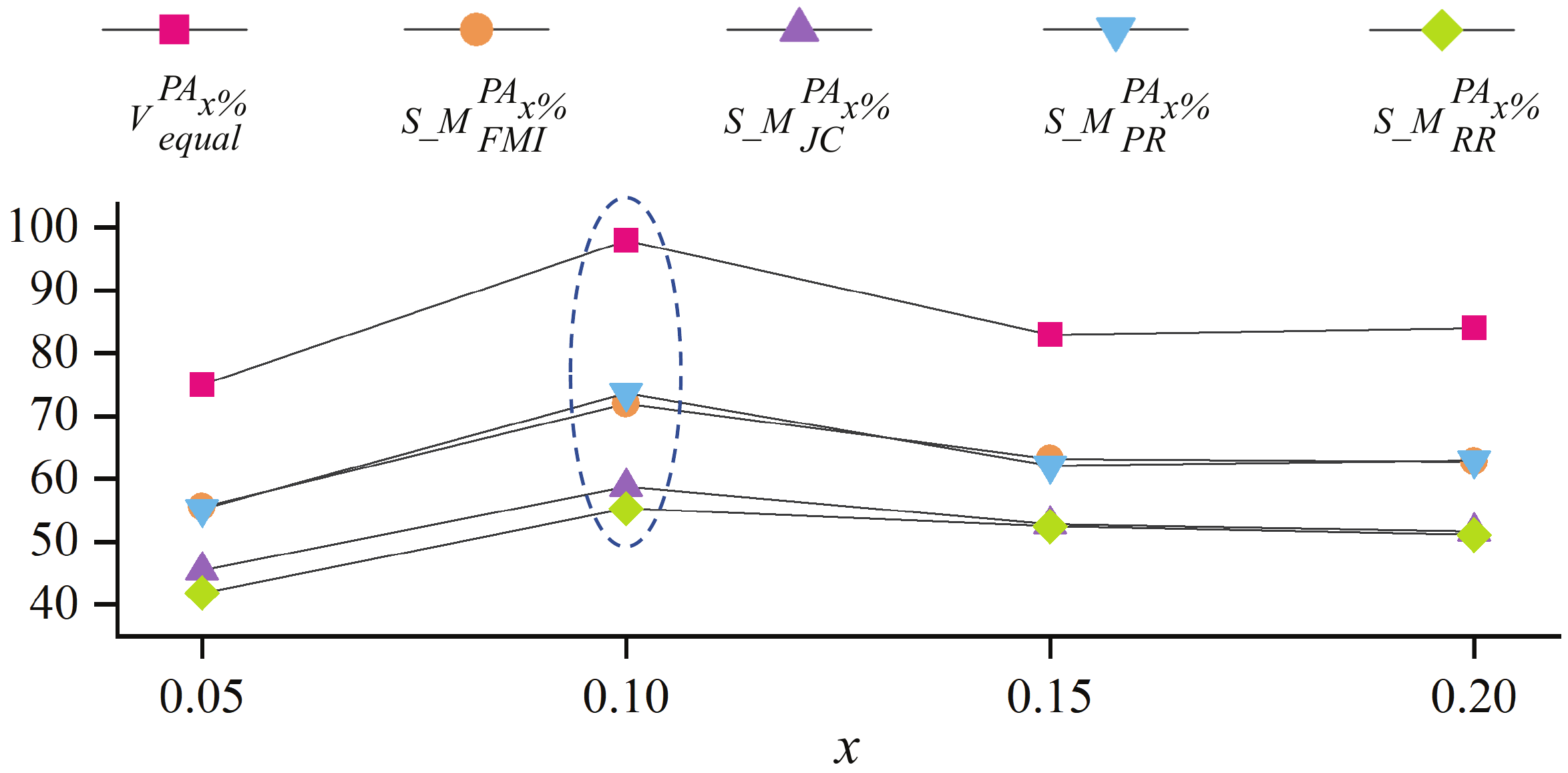} 
	\caption{Comparison in different values of \bm{$x$}} 
	\label{fig:rq1}  
\end{figure}

\vtwo{For those ``$k$ == $r$'' faulty versions, we use the four metrics introduced in Section~\ref{subsect:metrics} to evaluate the clustering effectiveness, as shown in Table~\ref{tab:rq1} and Figure~\ref{fig:rq1}. Taking ``$S\_M^{PA_{10\%}}_{FMI}$: 72.01'' as an example. It means that on 98 ``$k$ == $r$'' faulty versions  achieved by PA$_{10\%}$, the sum of the values of $FMI$, i.e., $FMI_1$ + $FMI_2$ + ... + $FMI_{98}$, is 72.01. We can observe that PA$_{10\%}$ delivers 72.01, 58.78, 73.62, and 55.32 points on the four metrics, which is more promising compared with that delivered by the other three variants.}

\vtwo{Based on these results, we can find that PA$_{10\%}$ performs best. We think that the Top-5\% riskiest statements are not sufficient for representing failures, while the Top-15\% and the Top-20\% riskiest statements may incur irrelevant information, which can negatively affect failure representation. Therefore, ``PA'' in the next two RQs  is ``PA$_{10\%}$'' in this RQ.}

\begin{table}[]\small
	\centering
	\caption{\label{tab:rq1} \tbdtwo{Comparison in different values of \bm{$x$}}} 
	\begin{tabular}{cccccc}
		\hline
		\rule{0pt}{8pt} \textbf{T} &  \bm{$V^T_{equal}$} &  \bm{${S\_M}^T_{FMI}$} &  \bm{${S\_M}^T_{JC}$} &  \bm{${S\_M}^T_{PR}$} &  \bm{${S\_M}^T_{RR}$} \\ \hline
		PA$_{5\%}$ & 75 & 55.57 & 45.45 & 55.23  &  41.85 \\
		PA$_{10\%}$ & \textbf{98} & \textbf{72.01} & \textbf{58.78} & \textbf{73.62}   & \textbf{55.32}  \\
		PA$_{15\%}$ & 83 & 63.23 & 52.85 & 62.11  &  52.46  \\
		PA$_{20\%}$ & 84 & 62.82 & 51.68 & 62.98  &  51.13  \\
		\hline         
	\end{tabular}
\end{table}

\subsection{RQ2: Impact analyses of components}
\label{subsect:rq2}
\vtwo{Revisiting Section~\ref{subsect:distance}, the distance metric of the proposed approach involves two components, i.e., the breakpoint level and the variable level. In this RQ, we further analyze the impact of each one. To that end, we compare the proposed approach with its two variants, PA$_{v^-}$ (keep only the breakpoint level and ablate the variable level) and PA$_{b^-}$ (keep only the variable level and ablate the breakpoint level) \sy{on SIR faulty versions}.}

\vtwo{As for PA$_{v^-}$, we do not consider original variable information at the breakpoints covered by both two failures, but only consider variables' name. Specifically, if two failures both cover a breakpoint, we simply calculate the value of $Distance^j_{var}$ in Formula~\ref{equ:dis_bpj} through dividing the intersection of the names of the variables collected by the two failures at the breakpoint by their union, rather than using Formula~\ref{equ:dis_var}. And as for PA$_{b^-}$, we merge variable information collected by a failure at all breakpoints into a hunk (without considering coverage of breakpoints), and feed such hunks of two failures into the variable level. That is, we directly use Formula~\ref{equ:dis_var} to measure the distance between two failures. The results are given in Figure~\ref{fig:rq2}.}

\subsubsection{The capability to estimating the number of faults of different components}
\label{subsubsect:rq2_1}
\begin{figure}  
	\includegraphics[height=2.6cm,width=8.5cm]{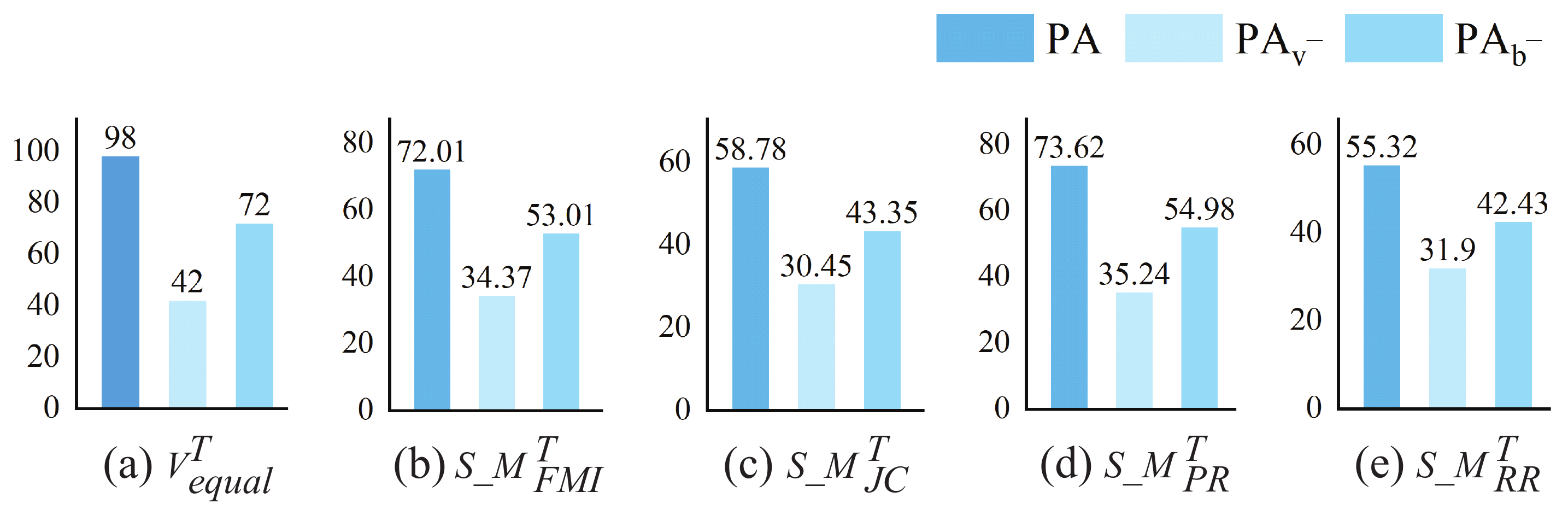} 
	\caption{Impact of components} 
	\label{fig:rq2}  
\end{figure}

\vtwo{Figure~\ref{fig:rq2}(a) compares the proposed approach with its two variants in terms of the capability to faults number estimation, exhibiting performance drops on the condition of an incomplete distance metric. Specifically, PA$_{v^-}$ and PA$_{b^-}$can correctly estimate the number of faults on 42 and 72 faulty versions, respectively, decreased by 57.14\% and 26.53\% respectively compared with the proposed approach.}

\subsubsection{The capability to clustering of different components}
\label{subsubsect:rq2_2}
\vtwo{The remaining four sub-figures in Figure~\ref{fig:rq2} perform the comparison in terms of the capability to clustering, from the perspectives of FMI, JC, PR, and RR. For example. the proposed approach can get 72.01 points on the metric FMI, while PA$_{v^-}$ and PA$_{b^-}$ get 34.37 and 53.01 points, respectively. With regard to the metrics JC, PR, and RR, we can observe a similar trend that PA$_{v^-}$ and PA$_{b^-}$ cause performance degradation.}

\vtwo{Based on these results, we can find that the default PA performs best, which demonstrates that each component in the distance metric positively contributes to the performance of the proposed approach. In particular, it is interesting that ablating the variable level harms the effectiveness of the proposed approach to a larger extent than ablating the breakpoint level, which double-confirms the intuition of this paper, i.e., the run-time values of program variables can be an effective failure distinguisher in failure indexing.}

\subsection{RQ3: Competitiveness of the proposed approach}
\label{subsect:rq3}

\begin{table}[]\small
	\centering
	\caption{\label{tab:rq3} \tbdtwo{Comparison with the state-of-the-art techniques}} 
	\begin{tabular}{lccccc}
		\hline
		\rule{0pt}{8pt} \qquad\textbf{T} &  \bm{$V^T_{equal}$} &  \bm{${S\_M}^T_{FMI}$} &  \bm{${S\_M}^T_{JC}$} &  \bm{${S\_M}^T_{PR}$} &  \bm{${S\_M}^T_{RR}$} \\ \hline
		(S) PA  & \textbf{98} & \textbf{72.01} & \textbf{58.78}  & \textbf{73.62} & \textbf{55.32} \\
		(S) MSeer & 68 & 51.63 & 42.50 &  53.67 & 31.75  \\
		(S) Cov$_{count}$ &48  & 37.83 & 31.47 & 37.71 & 26.97  \\
		(S) Cov$_{hit}$ & 34 & 26.61 & 22.35 & 23.83 &   15.53\\
		\rule{0pt}{9pt}(D) PA  & \textbf{37} & \textbf{36.96} & \textbf{36.93}  & \textbf{36.60} & \textbf{36.75} \\
		(D) MSeer & 29 & 29.00 & 29.00 & 29.00 &  29.00 \\ 
		(D) Cov$_{count}$ & 27 & 27.00&  27.00 & 27.00&  27.00\\
		(D) Cov$_{hit}$ & 27 & 27.00 &  27.00 &  27.00 & 27.00 \\
		\hline         
	\end{tabular}
\end{table}

\vtwo{As we mentioned in Section~\ref{sect:back}, SD-based and CC-based strategies are the most up-to-date and prevalent failure proximities to date. Therefore, we compare the proposed approach with these two for a robust and convincing evaluation. Specifically, as for the SD-based proximity, we select MSeer~\cite{gao2019mseer[8]} since it is the state-of-the-art in this class. And as for the CC-based proximity, we select Cov$_{hit}$~\cite{huang2013empirical[36], liu2008systematic[1]} since it is the most general configuration in this class. Moreover, \tbdtwo{considering that some works concern the impact of the execution frequency of program statements on debugging~\cite{shu2016fault[39], wen2012software[40]},} we create a variant of Cov$_{hit}$, i.e., Cov$_{count}$, which employs execution frequency rather than binary coverage as the fingerprinting function, as a baseline. To evaluate the proposed approach in a more comprehensive environment, in this RQ, we use both the simulated (SIR) and the real-world  (Defects4J) benchmarks. The results are given in Table~\ref{tab:rq3}, in which the first four rows (marked as ``(S)'') depict the results on SIR, while the last four rows (marked as ``(D)'') depict the results on Defects4J.}

\subsubsection{The capability to estimating the number of faults on SIR}
\label{subsubsect:rq3_1}

\vtwo{The proposed approach substantially outperforms all the baseline techniques regarding the capability of faults number estimation. Specifically, our approach can correctly predict the number of faults on 98 faulty versions on SIR, with 44.12\%, 104.17\%, and 188.24\% improvements compared with MSeer (68), Cov$_{count}$ (48), and Cov$_{hit}$ (34), respectively.}

\subsubsection{The capability to clustering on SIR}
\label{subsubsect:rq3_2}

\vtwo{The proposed approach consistently exceeds three baselines on all the four clustering metrics. For instance, if we focus on the comparison between our approach and MSeer, improvements are 39.47\%, 38.31\%, 37.17\%, and 74.24\%, regarding FMI, JC, PR, and RR, respectively. Considering the four metrics globally, the average improvement is 47.30\%. Similarly, in the contexts of comparing the proposed approach with Cov$_{count}$ and Cov$_{hit}$, the average improvements can be calculated as 94.37\% and 199.69\%, respectively.}

\subsubsection{The capability to estimating the number of faults on Defects4J} 
\label{subsubsect: rq3_3}
\vtwo{On all Defects4J faulty versions, the proposed approach can make $k$ equal to $r$ on 37 faulty versions, it is 27.59\%, 37.04\%, and 37.04\% higher than MSeer (29),  Cov$_{count}$ (27), and Cov$_{hit}$ (27), respectively.}

\subsubsection{The capability to clustering on Defects4J}
\label{subsubsect: rq3_4}
\vtwo{Similar to that in simulated scenarios, we can also observe that the proposed approach has a stronger capability of clustering than the baseline techniques in real-world scenarios. In particular, if we focus on the comparison between our approach and MSeer, improvements are 27.45\%, 27.34\%, 26.21\%, and 26.72\%, regarding FMI, JC, PR, and RR, respectively. Considering the four metrics globally, the average improvement is 26.93\%. Similarly, in the contexts of comparing our approach with Cov$_{count}$ and Cov$_{hit}$, the average improvements can be both calculated as 36.34\%.}

\vtwo{Based on these results, we can find that the proposed approach significantly outperforms the most up-to-date and prevalent techniques in the current field of failure indexing, in both simulated and real-world scenarios. Such promising outcomes demonstrate the competitiveness of our approach, and show the potential of the proposed program variable-based failure proximity.}

\begin{figure}  
	\centering  
	\includegraphics[height=3.4cm,width=8.5cm]{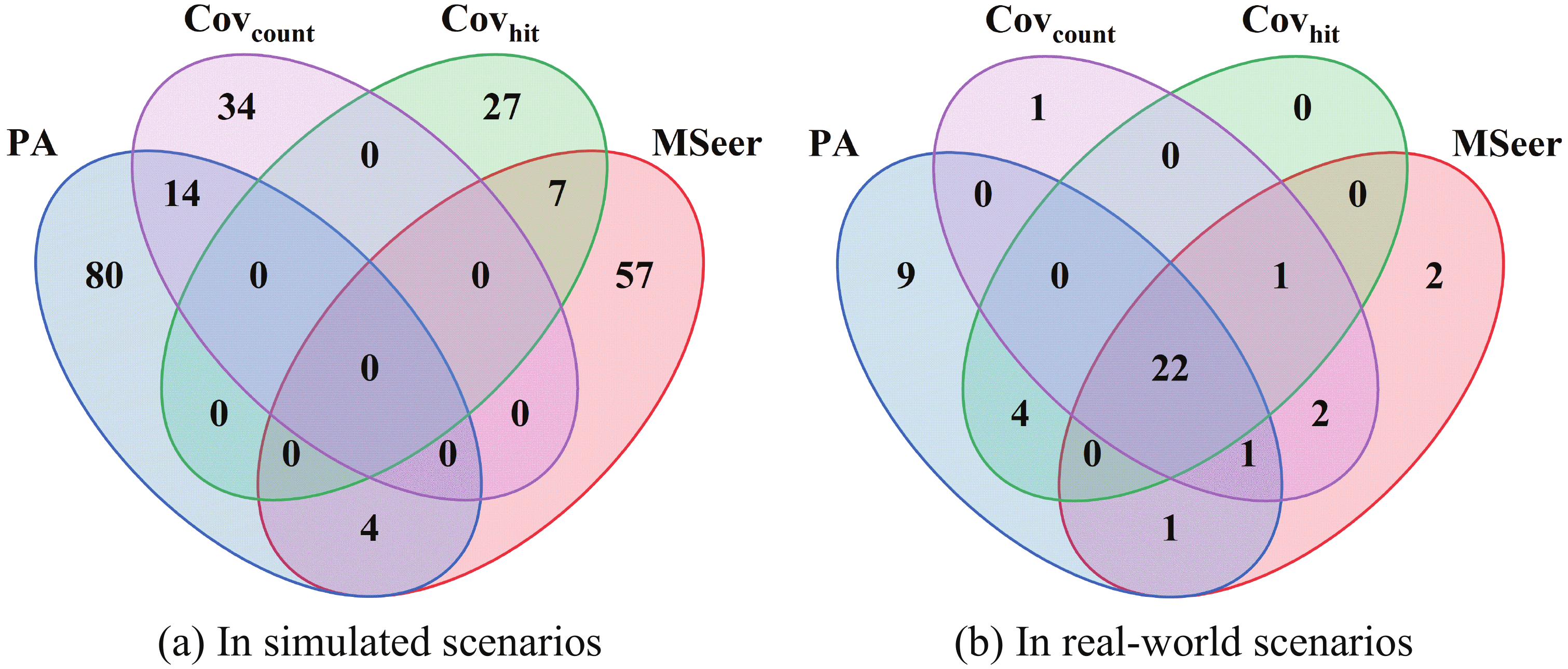} 
	\caption{The divergence of the ``\bm{$k$} == \bm{$r$}'' faulty versions} 
	\label{fig:venn}  
\end{figure}

\section{DISCUSSION}
\label{sect:discussion}

\subsection{Unique faulty versions handled by the proposed approach}
\label{subsect: generalization}

\vtwo{In the experiments, we find that those ``$k$ == $r$'' faulty versions achieved by different techniques are not exactly the same. In other words, some faulty versions can only be handled by a certain technique. We further discuss the proposed approach and the three baselines from this aspect, as shown in Figure~\ref{fig:venn}.}

\vtwo{In Figure~\ref{fig:venn}(a), we can find that of the 600 SIR faulty versions, 80 can only be handled by our proposed approach, 57, 34, and 27 can only be handled by MSeer, Cov$_{count}$, and Cov$_{hit}$, respectively. A similar observation can be drawn from Figure~\ref{fig:venn}(b): of the 100 Defects4J faulty versions, 9 can only be handled by our proposed approach, 2, 1, and 0 can only be handled by MSeer, Cov$_{count}$, and Cov$_{hit}$, respectively.} \tbdtwo{Though none of the failure indexing techniques is completely dominated by others, there are more faulty versions that can uniquely be handled by the proposed approach. This result further shows the competitiveness of our approach from a heuristic perspective, and indicates a potential future direction of combining the advantage of different failure proximities.}

\subsection{Efficiency of the proposed approach}
\label{subsect:efficiency}
\vtwo{The time costs of the proposed approach mainly involve three parts, i.e., the failure representation, the distance measurement, and the clustering. According to our investigation, the proposed approach typically spends 3.99 minutes and 5.90 minutes on average on generating the proxy for a failed test case, and spends 0.07s and 0.03s on average on measuring the distance between a pair of failed test cases, in simulated and real-world scenarios, respectively. After these two steps are ready, the clustering process typically takes only a few seconds. }

\section{THREATS TO VALIDITY}
\label{sect:threats}

\syrevised{Our experiments are subject to several threats to validity.}

\syrevised{The first is about the representativeness of the benchmark. We evaluate our approach on both simulated and real-world datasets. For the former, we adopt a diversity of mutation operators to inject faults, and for the latter, we collect projects from the industrial programming practice. Although this allows us to have higher confidence with respect to the generalization capability of the proposed approach, those benchmarks could still not be enough to represent different kinds of software systems. In the future, we plan to further evaluate our approach in larger-scale and more general environments.}

\syrevised{The second is about the choice of the evaluation metrics. We select four widely-used metrics, i.e., FMI, JC, PR, and RR, to quantitatively demonstrate the promise of our approach, but all of these four are external metrics, that is, the measurement of the clustering effectiveness is dependent on external information (i.e., the oracle groups). As another type of metrics, internal metrics are based on the generated clusters themselves, which could also contribute to our experimental evaluation, thus mitigating the bias incurred by only employing external metrics. In the future, we plan to integrate them into our work for a more robust evaluation.}

\section{RELATED WORK}
\label{sect:related}

\vtwo{As a very early work in this field, Podgurski et al. suggested using code coverage as a failure representer~\cite{podgurski2003automated[zxh42]}. Since then, such a CC-based failure proximity has been continuously adopted by stakeholders. For example, Huang et al. conducted an empirical study to investigate the impact of several factors on effectiveness of multi-fault localization, based on the CC-based failure proximity~\cite{huang2013empirical[36]}. Wu et al. clustered failures according to their coverage information using the Euclid distance, and prioritized them to facilitate the following fault localization~\cite{wu2020fatoc[zxh45]}.}

\vtwo{But the effectiveness of the CC-based failure proximity could be degraded if a fault triggers failures in different ways. To that end, Liu and Han regarded two failures as similar if they suggest roughly the same fault location. They introduced a statistical debugging tool~\cite{liu2005sober[zxh34]} to complete the mentioned suggestion process~\cite{liu2006failure[zxh3]}. Such an SD-based failure proximity attracts broad attention from academia. For example, Jones et al. produced a ranking list of suspiciousness using Tarantula~\cite{jones2005empirical[4]}, a well-known SBFL formula, to represent failures and index them based on the Jaccard distance~\cite{jones2007debugging[9]}. Gao and Wong utilized another SBFL formula, Crosstab~\cite{wong2011towards[zxh9]}, to deliver ranking lists that serve as the proxies for failures, and performed clustering using a revised Kendall tau distance~\cite{gao2019mseer[8]}.}

\vtwo{The motivation of SD-based failure proximities is to handle the scenario of the coverage of the failures having the same root cause is distinct. But if the coverage of the failures having different root causes is identical, neither of CC-based and SD-based tactics can work well. This paper utilizes the run-time values of program variables for getting rid of this bottleneck, showing a remarkable improvement.}

There are some recent works that introduce external profiles to support the failure indexing, such as code-independent features in regression testing~\cite{golagha2019failure[10]}, as well as code features and historical features in continuous integration~\cite{An2022[11]}. \vtwo{We do not consider such types of studies since they go beyond our research scope: 1) their source information cannot be always available, and 2) this paper focuses on failure indexing in the context of multi-fault localization.}

\section{CONCLUSION}
\label{sect:conclusion}
\syrevised{In this paper, we propose a novel type of failure proximity, namely, the program variable-based failure proximity, and further present a variable information-based failure indexing approach. The proposed approach mainly comprises the newly-defined fingerprinting function that integrates the run-time values of program variables to represent failures, and the distance metric designed to \vtwo{cooperate with} the fingerprinting function. Experiments demonstrate the competitiveness of the proposed approach. Specifically, compared with the state-of-the-art technique, our approach can achieve \vtwo{44.12\%} and \vtwo{27.59\%} improvements in faults number estimation, as well as \vtwo{47.30\%} and \vtwo{26.93\%} improvements in clustering effectiveness, in simulated and real-world environments, respectively. Besides, there are more faulty programs that can only be handled by the proposed approach compared with using the other techniques in our experiment.}

\syrevised{In the future, we plan to draw on deep learning methods to deliver a stronger failure indexing approach. A more comprehensive trial with larger and more general benchmarks as well as a broader spectrum of evaluation metrics is also being conceived.}

\bibliographystyle{ACM-Reference-Format}
\bibliography{ref}

\appendix

\end{document}